\pdfoutput=1
\documentclass{article}

\usepackage{listings}
\usepackage{epsfig}
\usepackage{url}
\usepackage{paralist}
\usepackage[utf8]{inputenc}
\usepackage[T1]{fontenc}
\usepackage{multirow}

\title{What Java Developers Know About Compatibility, And Why This Matters}

\author{
Jens Dietrich\\School of Engineering and Advanced Technology\\Massey University, Palmerston North, New Zealand\\ email:\url{j.b.dietrich@massey.ac.nz}\\ 
\and 
Kamil Jezek , Premek Brada \\
Faculty of Applied Sciences \\ University of West Bohemia, Pilsen, Czech Republic \\\url{{kjezek,brada}@ntis.zcu.cz}
}

\date{\today}

\begin{document}

\maketitle



\begin{abstract}

Real-world programs are neither monolithic nor static -- they are constructed 
using platform and third party libraries, and both programs and libraries 
continuously evolve in response to change pressure. In case of the Java 
language, rules defined in the Java Language and Java Virtual Machine 
Specifications define when library evolution is safe. These rules distinguish 
between three types of compatibility - binary, source and behavioural. We claim 
that some of these rules are counter intuitive and not well-understood by many 
developers. 
We present the results of a survey where we quizzed developers about their
understanding of the various types of compatibility. 414
developers responded to our survey. We find that while most programmers are
familiar with the rules of source compatibility, they generally lack knowledge
about the rules of binary and behavioural compatibility. This can be problematic
when organisations switch from integration builds to technologies that require
dynamic linking, such as OSGi.  We have assessed the gravity of the problem by
studying how often linkage-related problems are referenced in issue tracking
systems, and find that they are common.

\end{abstract}

\section{Introduction}

Modern programming languages like Java support dynamic linking where a client program can be 
executed with libraries that have been compiled separately. While the libraries 
used at runtime are usually also present when the program is compiled, it is not 
required that the \textit{same versions} of these libraries are used for 
compilation. This addresses some important use cases, such as the deployment of 
newer versions of a library with improvements such as bug fixes or 
better performance. As long as the APIs (application programming interfaces) defined in these libraries don't 
change, this works well. Unfortunately, APIs do change when libraries evolve 
\cite{DigJohnson:06,cossette2012seeking,brokenpromises}. When this happens, 
programmers are suddenly confronted with different sets of rules 
\cite{KindsOfCompatibility}: the rules of \textit{source compatibility} are used 
by the compiler when a program is compiled against a library, while the rules of 
\textit{binary compatibility} are used when a program is linked against a 
library that has been compiled separately. To make things even more complicated, the 
Java Language Specification defines binary compatibility strictly with respect 
to linking \cite[ch. 13]{JL7Spec}. This does not cover all problems that can occur when the 
respective program is executed with a particular library in the classpath. This 
leads to a third type of compatibility, \textit{behavioural compatibility}.  

To illustrate the different types of compatibility, consider the following 
examples in listing \ref{exampleRC1} and \ref{exampleEXC1}. Both examples 
consist of a class \texttt{Main} that is compiled with the first version of a 
class \texttt{Foo} in a library \texttt{ lib-1.0.jar}, and then executed with 
another version of \texttt{Foo} in a library \texttt{lib-2.0.jar}.

\begin{lstlisting}[label=exampleRC1,caption=Specialising the return type of a method]
// lib-1.0.jar
package lib.specialiseReturnType1;
public class Foo {
	public static java.util.Collection getColl() {
		return new java.util.ArrayList();
	}
}
// lib-2.0.jar
package lib.specialiseReturnType1;
public class Foo {
	public static java.util.List getColl() {
		return new java.util.ArrayList();
	}
}
// client program
package specialiseReturnType1;
import lib.specialiseReturnType1.Foo;
public class Main {
	public static void main(String[] args) {
		java.util.Collection coll = Foo.getColl();
		System.out.println(coll);
	}
}
\end{lstlisting}

These two examples demonstrate that the different types of compatibility are 
inconsistent and not intuitive. In the first example, the return type of a used 
method is specialised. While this is usually source compatible\footnote{The 
exception is when the method is overridden}, it is not binary compatible as the 
linker does not use subtype reasoning when a method reference is resolved. In 
this case, a \texttt{NoSuchMethodError}  is thrown. 
In other words, the only required correction
is recompilation of the program with version 2.0 of the library.
No source code modification is needed.

\begin{lstlisting}[label=exampleEXC1,caption=Adding a checked exception to a method]
// lib-1.0.jar
package lib.exceptions2;
public class Foo {
	public static void foo() {}
}
// lib-2.0.jar
package lib.exceptions2;
import java.io.IOException;
public class Foo {
	public static void foo() throws IOException {
		throw new IOException();
	}
}
// client program
package exceptions2;
public class Main {
	public static void main(String[] args) {
		lib.exceptions2.Foo.foo();
	}
}
\end{lstlisting}

The second example shows that while information about the exceptions thrown is
present in the byte code as part of the signature \cite[ch. 
4.3.4]{JVM7Spec}, and can even be queried through the reflection API, this 
information is not part of the descriptor \cite[ch. 4.3.3]{JVM7Spec} used for 
linking. I.e., the uncaught checked exception is not detected through static 
analysis during linking, but only at runtime when the exception is actually 
thrown when \texttt{foo()} is invoked, and the program exits with an error. This 
is therefore an example of behavioural incompatibility, although it would be 
more intuitive if this was binary incompatible.

The mismatch between the different notions of compatibility has increased as the 
Java language has evolved. The development of Java language was driven by 
programmer productivity, while preserving binary compatibility  
\cite{JDKCompatibilityRegions}. Features like nested and inner classes, generic 
types and auto boxing / unboxing have added simplicity and expressiveness to the 
language with minimal or no changes to the byte code format. In case of generic 
types, this has lead to erasure. But even 
changing a field type from a primitive type to its wrapper type or vice versa 
breaks binary compatibility (listing \ref{exampleBOX1}), even though this 
problem can easily be solved by recompilation as the compiler applies auto 
boxing or unboxing, respectively.

\begin{lstlisting}[label=exampleBOX1,caption=Wrapping a primitive field]
// lib-1.0.jar
package lib.primwrap1; 
public class Foo {
	public static int MAGIC = 42; 
}
// lib-2.0.jar
package lib.primwrap1; 
public class Foo {
	public static Integer MAGIC = new Integer(42);
}
// client program
package primwrap1; 
import lib.primwrap1.Foo;
public class Main {
	public static void main(String[] args) { 
		int i = Foo.MAGIC; 
		System.out.println(i);
	}
}
\end{lstlisting}

It is reasonable to ask here why this matters. In many cases, a project is built 
(compiled and unit tested) against the very libraries it uses at runtime, and 
widely used build and continuous integration tools facilitate this approach. 
This circumvents the problem. However, if libraries are individually upgraded, 
problems start to occur. The JLS envisages this approach: ``Development tools 
for the Java programming language should support automatic recompilation as 
necessary whenever source code is available. Particular implementations may also 
store the source and binary of types in a versioning database and implement a 
ClassLoader that uses integrity mechanisms of the database to prevent linkage 
errors by providing binary-compatible versions of types to clients'' \cite[ch. 
13]{JL7Spec}. While preserving binary compatibility is a major objective for the 
JRE \cite{JDKCompatibilityRegions}, this is not the case for many other 
libraries. In our previous work \cite{brokenpromises} we have shown that binary 
compatibility is often broken when open source libraries evolve. In particular, 
this applies to popular libraries such as ANT, ANTLR, Hibernate and Lucene. The 
major reason that this matter is, however, the popularity of OSGi 
\cite{OSGi} -- a framework that heavily relies on dynamic linking to 
support a flexible lifecycle for components in long-running, high-availability 
applications. The problem might be further accelerated when modularity support is added 
to the Java platform (project Jigsaw), although it is not clear at this point whether Jigsaw will support dynamic modules \cite{Jigsaw2}.

The rest of this paper is organised as follows. In section \ref{SurveyDesign}, 
we discuss the design of the survey we conducted to find out what developers 
know about the different types of compatibility. We analyse the responses in 
section \ref{Responses}. In section \ref{ImpactAnalysis}, we try to assess the 
size of the problem in practice by analysing 
issue tracking systems for references to errors caused by linkage related 
problems. In section \ref{Threats} we discuss several treats to validity.
We finish the paper with a conclusion, including a discussion of 
related and future work.

\section{Survey Design}
\label{SurveyDesign}

The survey is based on a set of Java library evolution puzzlers we have 
developed for 
training\footnote{\url{
http://www.slideshare.net/JensDietrich/presentation-30367644}}, in a style 
inspired by \cite{JavaPuzzlers}. The full survey contains 7 questions about
the 
background of the respondent, and 25 puzzlers -- 21 standard puzzlers with 2 
questions each, and 4 constant inlining puzzlers with 1 question per puzzler -- 
a total of 46 technical questions. A few days after the survey had opened we 
realised  that many respondents only answered the first few questions, and then 
abandoned the survey. We therefore created a second, 
shorter survey with the same set of 7 background questions, 9 standard and 4 
constant inlining puzzlers, a total of 22 technical questions. The question in 
the short survey are a subset of the questions in the full survey.

\subsection{Respondent Background Questions}

We have asked respondents the following set of background questions to assess 
their level of relevant experience.

\begin{enumerate}
\item Which programming languages do you regularly use? 
Choices were: Java, C, C++, Python, Ruby, ObjectiveC, C\#, JavaScript, other 
JVM-based languages (Scala, Groovy, etc), other.
\item How many years of Java programming experience do you have? 
Choices were: less than 1 year, 1-3 years, 4-10 years, more than 10 years.
\item How would you rank your Java expertise?
Choices were: novice, some experience, knowledgeable, expert, guru.
\item Are you familiar with the following Java concepts? Only answer this
question if you chose novice as an answer to the previous question.
Choices were: interfaces, inheritance, the difference between source code and 
byte code, the difference between errors, checked and runtime exceptions, 
wrapper types (boxing/unboxing), generic parameter types, the classpath.
\item Which of the following technologies have you used?
Choices were: Spring, OSGi, OSGi extensions (Eclipse plugins, Spring DM, etc), 
WebStart, J2EE application servers, ASM, BCEL or other byte code analysis tools, 
AspectJ or other AOP tools, Antlr or other parser generators, Ant, Maven or 
other build tools, Jenkins, Hudson or other continuous integration tools.
\item Have you designed frameworks, libraries or APIs? 
This was a yes/no question.
\item What is your main occupation?
Choices were: student, academic, industry research, programmer, project manager.

\end{enumerate}

\subsection{Technical Questions Overview}

The technical questions are summarised in table \ref{tab:question-overview}. 
Each question in the table belongs to a category and the question's puzzler is
implemented as a simple program. Puzzlers are split into packages, the respective package names 
are shown in the third table column. Given a package name \texttt{<pck>}, each puzzler  consists of three classes 
named as follows:

\begin{enumerate}
	\item a class \texttt{lib.<pck>.Foo}  packaged in a library 
\texttt{lib-1.0.jar}
	\item a (modified) class \texttt{lib.<pck>.Foo}  packaged in a library 
\texttt{lib-2.0.jar}
	\item a class \texttt{<pck>.Main} with a \texttt{main} method that uses \texttt{lib.<pck>.Foo}
\end{enumerate}

The code and an ANT script to execute the experiments is available from the following public code repository:  \\
\\
\texttt{\url{https://bitbucket.org/jensdietrich/java-library-evolution-puzzlers}}
\\

The source code for the questions is in the \texttt{/examples} folder and its respective sub folders for both versions of the library and the client program. 
The root folder contains an ANT script that can be used to compile and execute the respective scenario. The correct answers are defined with respect to the output of this script, confirmed by cross-referencing the output with the Java Language Specification \cite{JL7Spec}. Given a unique package name \texttt{<pck>}, the following command can then be used to run the compilation and linking script for this puzzler / question:\\

\texttt{ant -Dpackage=pck}
\\

There are two types of problems. All problems except the problems in the constant inlining category are \textit{standard puzzlers}. For each standard puzzler, we have asked the following two questions: 

\begin{enumerate}
	\item Can the version of the client program compiled with \texttt{lib-1.0.jar} be executed with \texttt{lib-2.0.jar} ? The three possible answers are:
	\begin{enumerate} 
		\item no, an error occurs
		\item yes, but the behaviour of the program changes
		\item yes, and the behaviour of the program does not change
	\end{enumerate}
	\item Can the client program be compiled and then executed with \texttt{lib-2.0.jar} ? The three possible answers are:
	\begin{enumerate} 
		\item no, compilation fails
		\item yes, but the behaviour of the program is different from the program version compiled and executed with \texttt{lib-1.0.jar}
		\item yes, and the behaviour of the program is the same as the program version compiled and executed with \texttt{lib-1.0.jar}
	\end{enumerate}
\end{enumerate}

Note that we avoided references to the standard definitions of binary and source
compatibility in the specification documents on purpose, as we can not assume
that the majority of developers is familiar with these documents. We defined
a behaviour change as follows in the survey: ``a behaviour change is either a
different console output or a situation where the execution of one program
version throws an exception, but the the execution of the other program version does not''.

Table \ref{tab:question-overview} uses wildcards in the question ids. The wildcard is replaced by ``-UPGR'' (dynamic upgrade) for the first question, and by ``-RECP'' (recompile and upgrade) for the second question. For the constant inlining category, only one question is asked. This will be explained below. We refer to these three question types as UPGR, RECP and INL, respectively. 

\begin{table}[!h]
\caption{Question Overview}
\label{tab:question-overview}
\begin{tabular}{ |p{3.2cm}p{2.3cm}p{3.2cm}p{1.8cm}| }
\hline
category & question id & package name & short survey\\
\hline \hline
	interfaces & IF1-* & addtointerface & yes \\
	interfaces & IF2-* & removefrominterface1 & no \\
	interfaces & IF3-* & removefrominterface2 & no \\
	interfaces & IF4-* & removefrominterface3 & no \\
	\hline
	method descriptors & SP\_RET1-* & specialiseReturnType1 & yes \\
	method descriptors & SP\_RET2-* & specialiseReturnType2 & no \\
	method descriptors & SP\_RET3-* & specialiseReturnType3 & no \\
	method descriptors & SP\_RET4-* & specialiseReturnType4 & no \\
	method descriptors & GEN\_PAR1-* & generaliseParamType1 & yes \\
	method descriptors & GEN\_PAR2-* & generaliseParamType2 & no \\
	method descriptors & GEN\_PAR3-* & generaliseParamType3 & yes \\
	\hline
	exceptions & ADD\_EXC1-* & exceptions1 & no \\
	exceptions & ADD\_EXC2-* & exceptions2 & yes \\
	exceptions & SPEC\_EXC1-* & exceptions4 & no \\
	exceptions & REM\_EXC1-* & exceptions5 & yes \\
	exceptions & REM\_EXC2-* & exceptions6 & no \\
	\hline
	auto (un)boxing & BOX1-* & primwrap1 & yes \\
	auto (un)boxing & BOX2-* & primwrap2 & no \\
	\hline
	generic param types & GEN1-* & generics1 & yes \\
	generic param types & GEN2-* & generics2 & yes \\
	\hline
	constant inlining & CON1 & constants1 & yes \\
	constant inlining & CON2 & constants2 & yes \\
	constant inlining & CON3 & constants3 & yes \\
	constant inlining & CON4 & constants4 & yes \\
	\hline
	others & NEST-* & ghost & no \\
\hline
\end{tabular}
\end{table}

\subsection{Interface Puzzlers}

The puzzlers in this category describe problems that occur when methods are added to or removed from interfaces implemented by a client program. In IF1, a method is added to the interface that is not actually being used by the class implementing the interface. This is binary compatible, but not source compatible. To solve the other problems in this category, respondents also had to understand the \texttt{@Override} annotation (IF2, IF3)  and Java method lookup (IF4). 

\subsection{Method Descriptor Puzzlers}

The descriptor of a method is the combination of parameters plus the return
type, and this information is used at runtime for linking and method dispatch
\cite[ch. 15.12.2]{JL7Spec}. When refactoring methods, specialising return types
and generalising parameter types is generally safe. This can be seen as 
special cases of strengthening post-conditions and weakening pre-conditions, respectively.

However, in both cases the method descriptors are changed, causing binary incompatibility. More precisely, this results in an instance of \texttt{NoSuchMethodError} being thrown. SP\_RET1 is shown in listing \ref{exampleRC1}. The other questions in this category are variants of this question, also including the narrowing of primitive return types (SP\_RET2) and the (unsafe) widening of primitive parameter types (GEN\_PAR3). GEN\_PAR2 is a scenario where generalising a parameter type leads to ambiguity and compilation fails. 

\subsection{Exception Puzzlers}

Exceptions (including checked exceptions) are not part of the descriptor. This 
means that changes to the exceptions declared by a method like adding or 
generalising are generally binary compatible but often behavioural incompatible.
This 
is the case in example ADD\_EXC2 (listing  \ref{exampleEXC1}). ADD\_EXC1 is a 
similar scenario using a runtime exception. On the other hand, certain changes 
to declared exceptions that seem uncritical like removing a declared exception 
can lead to source incompatibilty as the compiler detects that catch blocks 
become unreachable \cite[ch. 14.21]{JL7Spec}. Question REM\_EXC1 is based on 
such a scenario. Question REM\_EXC2 is noteworthy  as the behaviour of the 
current Java (OpenJDK SE Runtime Environment 1.7.0\_45-b18) implementation differs from the specification due to 
a bug in the JLS. This bug was reported and accepted\footnote{Email 
conversation with Alex Buckley, 11 October 2013}, and we defined the correct 
answer for this question with respect to the behaviour of the  Java 7 
implementation used.

\subsection{Auto (Un)Boxing Puzzlers}

Auto boxing / unboxing were introduced in Java 5.0. This feature hides the
differences between object (wrapper) types and their corresponding primitive
types. However, on the byte code level they are still treated as completely
different types. In BOX1 (listing \ref{exampleBOX1}), the library declares a
public field of type \texttt{int} that is read by the client to assign a value
to a variable declared as \texttt{int}. The type of this field is then changed
to \texttt{java.lang.Integer}. This is not binary compatible but source
compatible as the compiler can apply auto unboxing. More precisely, this results
in an instance of \texttt{NoSuchFieldError} to be thrown. BOX2 reverses this
scenario -- the field type is changed from \texttt{java.lang.Integer} to
\texttt{int}. The result is the same.

\subsection{Generic Type Puzzlers}
\label{puzzler:gen}
Generic types were also introduced in Java 5.0.  Erasure \cite[ch. 4.6]{JL7Spec} is used to achieve binary compatibility so that existing non-generic clients can use generic code. The questions in this category describe two  situations resulting from this. In the first puzzler (GEN1) the library defines a method \texttt{List<String> getList()} that returns a list containing \texttt{``42''}. This method is then changed to a method \texttt{List<Integer> getList()} that returns a list containing \texttt{42}. The client accesses the list, assigns it to a variable \texttt{List<String> list } and prints its size to the console. Due to erasure, this is binary compatible, but not source compatible as the compiler detects that the assignment is illegal. The second puzzler (GEN2) is very similar, but this time the client loops over the list and prints its element to the console. This is still binary compatible but fails with a class cast exception due to the \texttt{checkcast} statements generated by the compiler when the elements of the list are accessed. This is therefore an example of behavioural incompatibility.

\subsection{Constant Inlining Puzzlers}

The structure of the four questions in this category is slightly different. Listing \ref{exampleCONST1} shows the code used in CON1. 

\begin{lstlisting}[label=exampleCONST1,caption=Constant Inlining]
// lib-1.0.jar
package lib.constants1; 
public class Foo {
   public static final int MAGIC = 42; 
}
// lib-2.0.jar
package lib.constants1;
public class Foo {
   public static final int MAGIC = 43;
}
// client program
package constants1;
import lib.constants1.*;
public class Main {
   public static void main(String[] args) {
      System.out.println(Foo.MAGIC);
   }
}
\end{lstlisting}

There is only one question for each problem in this group: whether 42 or 43 is
printed to the console when the program compiled with version 1.0 of the library
is executed with version 2.0 of this library. CON2 uses strings instead of
integers, CON3 defines an integer constant using an expression instead of a
literal, and in CON4 the wrapper type \texttt{java.lang.Integer} is used instead
of \texttt{int}. It turns out that constants of type \texttt{int} and
\texttt{String}  are inlined, and that the compiler uses constant folding to evaluate expressions. This
means that 42 is printed when the programs in CON1, CON2 or CON3 are executed.
However, \texttt{Integer} values are not inlined, and 43 is printed in case of
CON4. 

\subsection{Other Puzzlers}

NEST is a scenario where the client program uses a static method defined in a class \texttt{lib.ghost.Foo.Bar}. In the first version, \texttt{Bar} is a static nested class within \texttt{lib.ghost.Foo}, while in the second version \texttt{Bar} is refactored to a top-level class within the package \texttt{lib.ghost.Foo}. This is binary incompatible but source compatible as the byte code represention changes from \texttt{lib/ghost/Foo\$Bar} to \texttt{lib/ghost/Foo/Bar} while the source code representation (\texttt{lib.ghost.Foo.Bar}) remains the same. 

\section{Responses}
\label{Responses}

\subsection{Access to Raw Survey Data}

The raw data exported from SurveyMonkey can be accessed in the following repository:  \\
\\
\texttt{\url{https://bitbucket.org/jensdietrich/java-library-evolution-puzzlers}}
\\ \\
The data is available as a set of Excel files in the  \texttt{/survey/results/rawdata/} folder.

\subsection{Overview}

The survey was open between 15 November and 31 December 2013. During this 
period, 184 respondents started the short version of the survey, while 241 
respondents started the full version. We asked respondents doing the full
survey 
whether they had already completed the short survey. Only 11 respondents 
answered yes. Assuming that everybody answered this question correctly, and 
nobody completed the short survey after the full survey, this gave us 414 
unique respondents. To avoid double counting, we removed all answers for 
questions that are in both surveys for responses by people who indicated that 
they had completed both surveys.  
 
 The number of valid answers for the technical questions ranged from 49 (REM\_EXC2-RECP) to 295 (IF1-UPGR and IF1-RECP). 
 
 \subsection{Background of Respondents}
 
\begin{figure}[!h]
	\centering
	\includegraphics[width=0.8\textwidth]{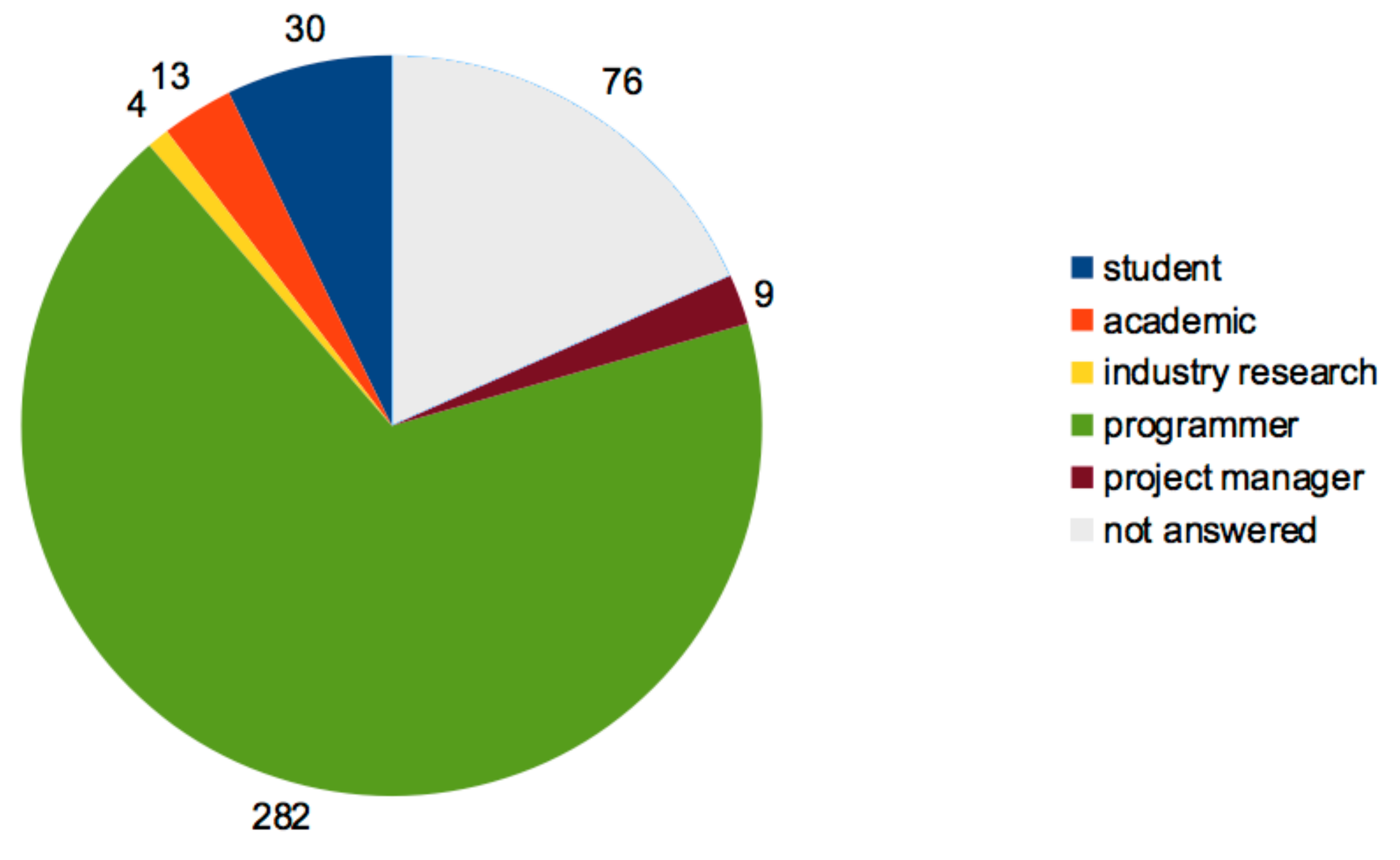}
	\caption{Occupation of survey respondents}
	\label{fig:occupation}
\end{figure}

\begin{figure}[!h]
	\centering
	\includegraphics[width=0.8\textwidth]{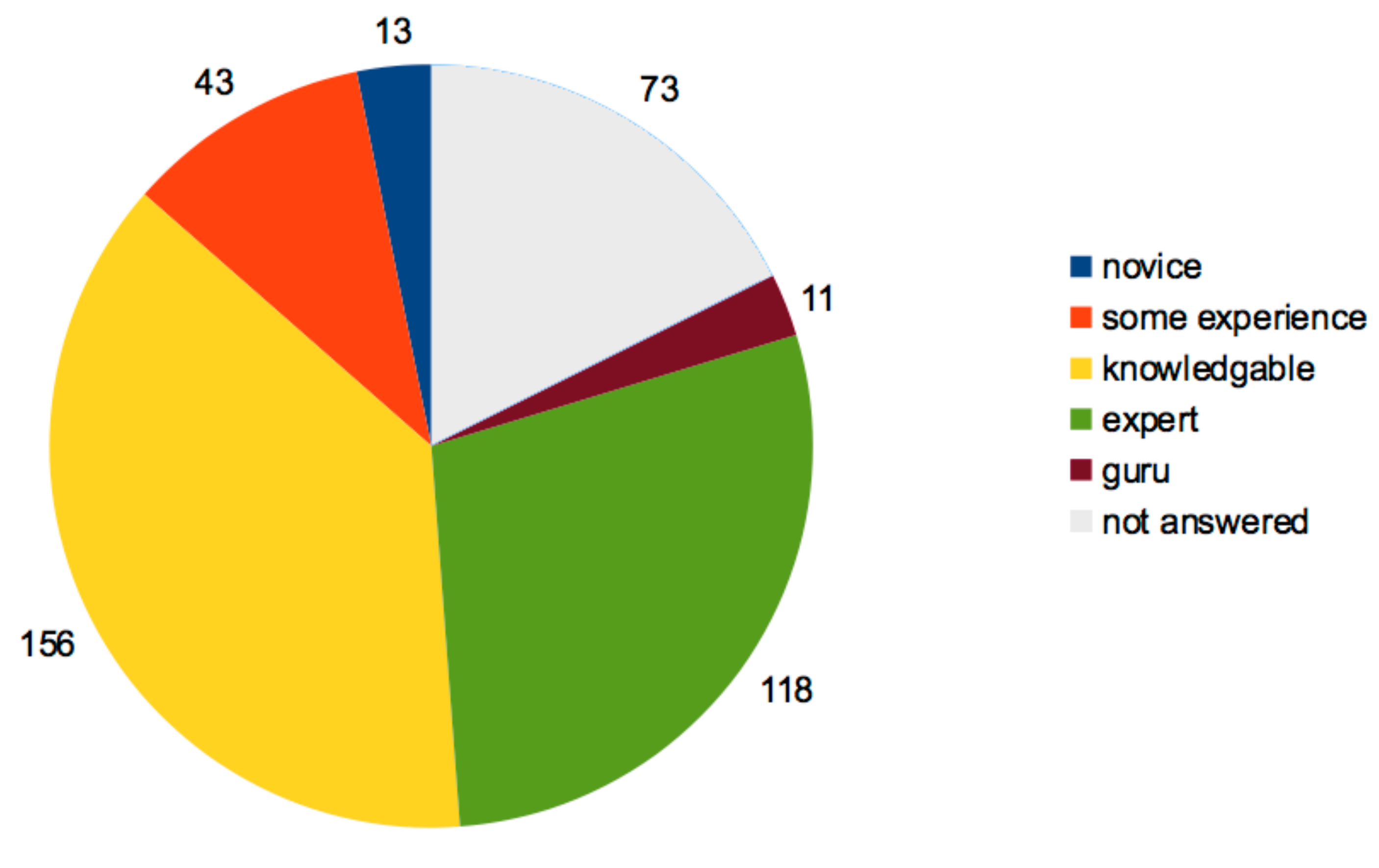}
	\caption{Self-assessed level of Java experience of survey respondents}
	\label{fig:levelofexperience}
\end{figure}

\begin{figure}[!h]
	\centering
	\includegraphics[width=0.8\textwidth]{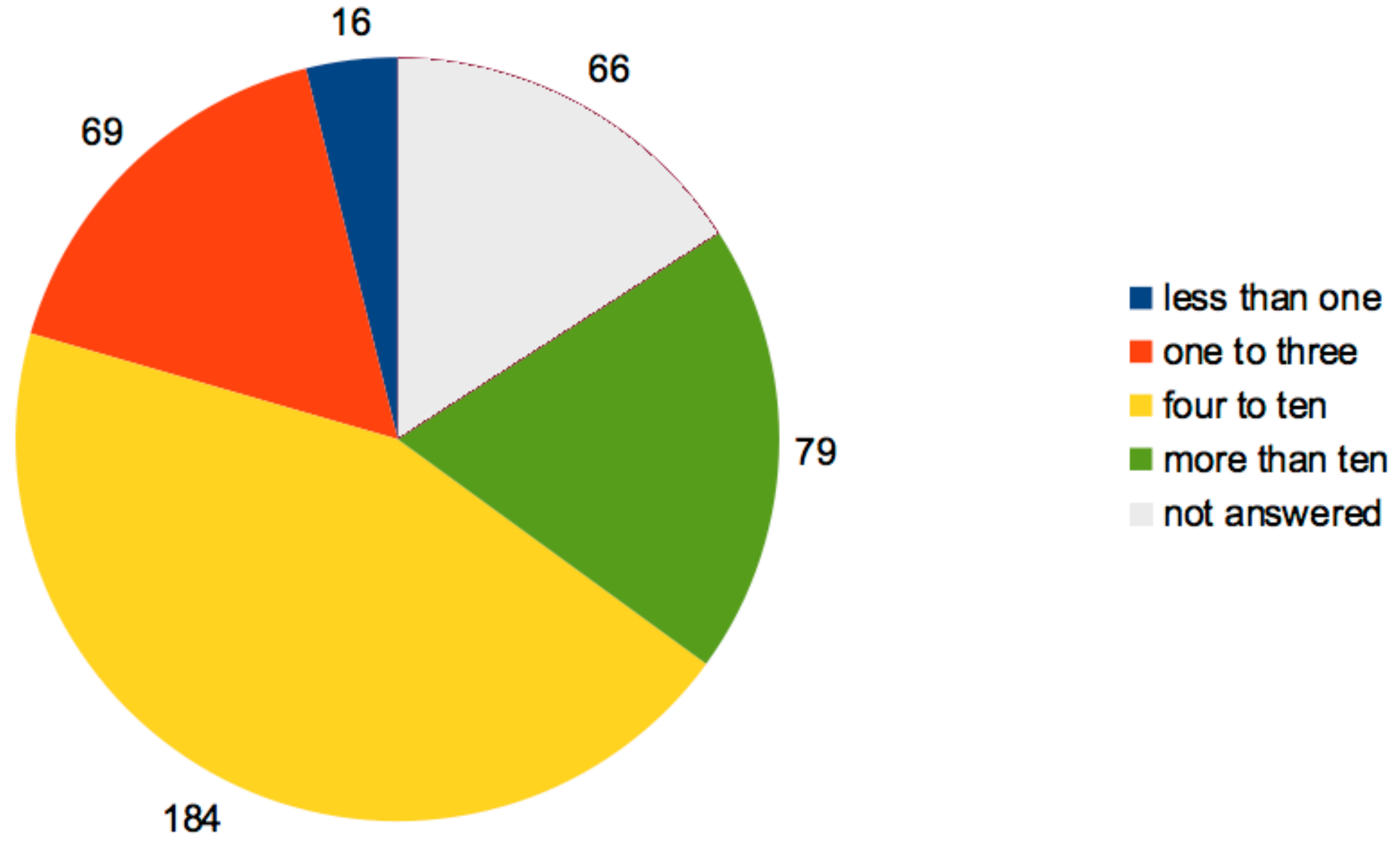}
	\caption{Years of Java experience of survey respondents}
	\label{fig:yearsofexperience}
\end{figure}

The vast majority of respondents are programmers (figure \ref{fig:occupation}), have 4 or more years of experience with Java technologies  (figure \ref{fig:yearsofexperience}) and self-assess their familiarity with Java technology as either knowledgeable or expert (figure \ref{fig:levelofexperience}).  


\subsection{Analysis}

Of all the answers provided by respondents, only 62\% were correct. The percentage of correct answers is much better for the RECP questions
 (76\%) than for the UPGR (51\%) and INL (52\%) questions. This indicates that respondents are more familiar with the rules of source compatibility than with the rules of binary and behavioural compatibility. 
 
\begin{figure}[tbh]
	\centering
	\includegraphics[width=1.0\textwidth]{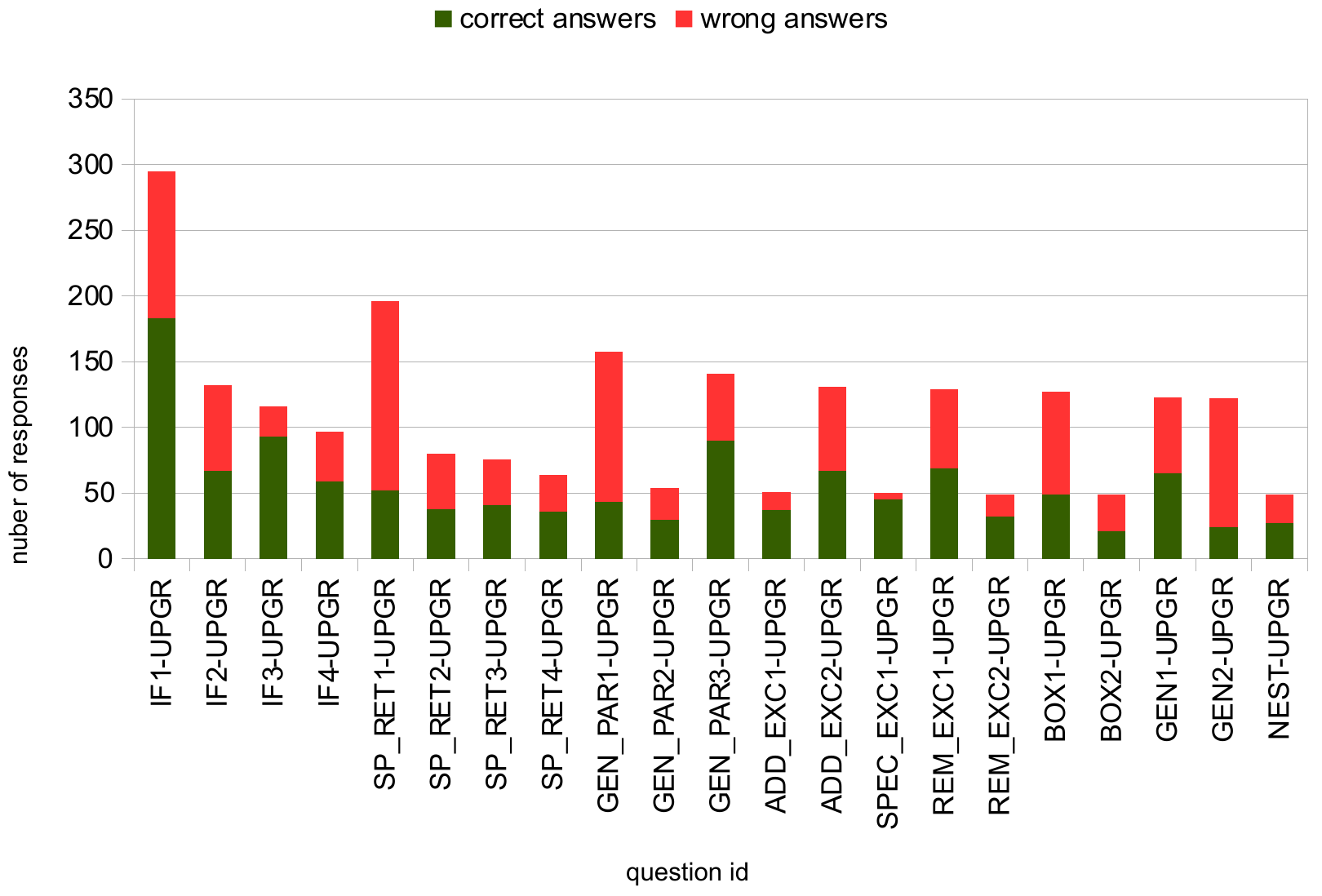}
	\caption{Overview -- answers for UPGR (library
upgrade) questions}
	\label{fig:bincomp-overview}
\end{figure}

\begin{figure}[!h]
 	\centering
 	\includegraphics[width=1.0\textwidth]{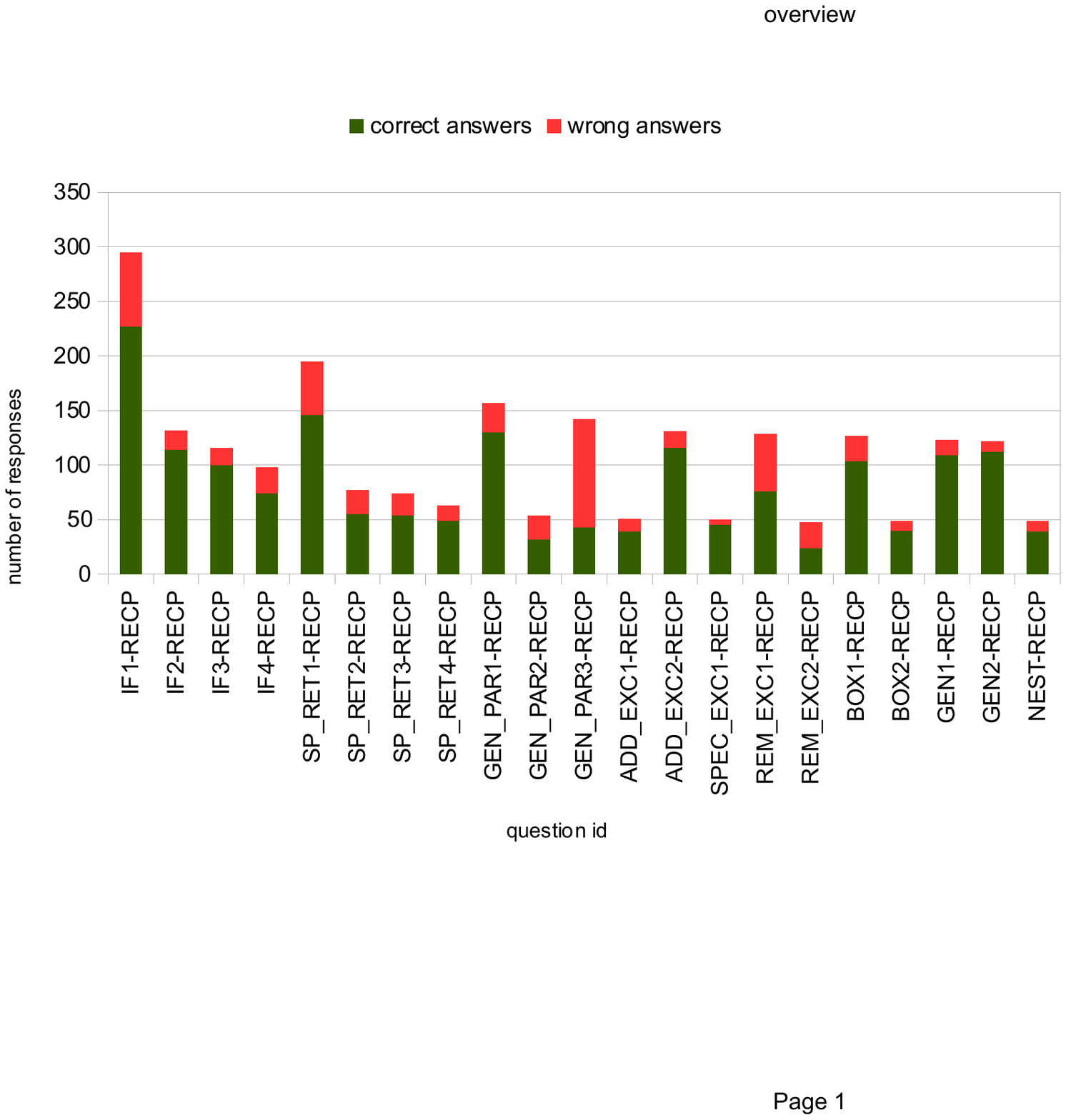}
 	\caption{Overview -- answers for RECP (recompilation)
questions}
 	\label{fig:srccomp-overview}
 \end{figure}

\begin{figure}[!h]
  	\centering
  	\includegraphics[width=1.0\textwidth]{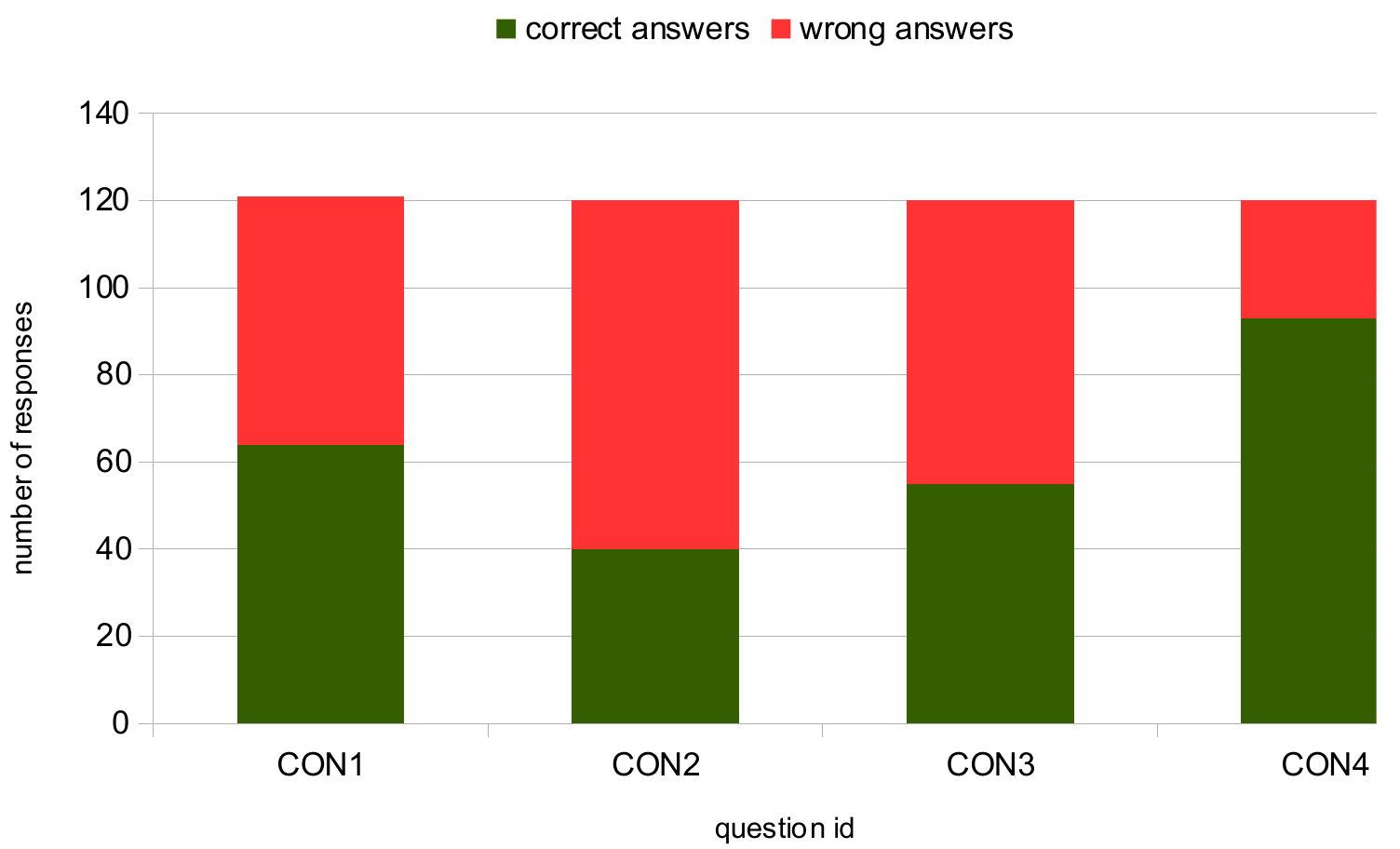}
  	\caption{Overview -- answers for constant inlining puzzler questions}
  	\label{fig:inline-overview}
  \end{figure}

 Figures \ref{fig:bincomp-overview}, \ref{fig:srccomp-overview} and \ref{fig:inline-overview} show the total numbers of correct and incorrect answers for 
 the questions in the UPGR, RECP and INL categories, respectively. The spikes in the total number of responses represent questions that were part of the short survey.  In general, there are more wrong answers for the questions that were also part of the short survey. This is due to the fact that more experienced developers completed the full survey. 
 
 The correct answer ratio is particularly low for the first motivational example
discussed in the introduction: only 27\% of respondents answered SP\_RET1-UPGR
(listings \ref{exampleRC1}) correctly. Other questions with a high percentage of
wrong answers are GEN\_PAR1-UPGR (27\% correct, generalising a parameter type is
binary incompatible), and GEN2-UPGR (20\%). In the case of GEN2-UPGR (see also
section \ref{puzzler:gen}), many respondents answered that the upgraded library
is binary incompatible. However, it is only behavioural incompatible -- no
linkage error is generated, but a runtime exception
(\texttt{ClassCastException}) is thrown when the program is executed.
 
 While most respondents answered the RECP type questions correctly, only 30\% got GEN\_PAR3-RECP (listing \ref{genParamType3}) right. This question is inspired by the classical Java puzzlers \cite{JavaPuzzlers}, where compilation succeeds but the behaviour of the program is changed by the recompilation due to the loss of precision when the widening conversion from \texttt{int} to \texttt{float} is performed \cite[ch. 5.1.2]{JL7Spec}. 
 
 \begin{lstlisting}[label=genParamType3,caption=Generalising a Parameter Type]
 // lib-1.0.jar
package lib.generaliseParamType3;
public class Foo {
	public static boolean isEven(int i) {
		return i%2==0;
	}
}
 // lib-2.0.jar
package lib.generaliseParamType3;
public class Foo {
	public static boolean isEven(float i) {
		return i%2==0;
	}
}
 // client program
package generaliseParamType3;
import lib.generaliseParamType3.Foo;
public class Main {
	public static void main(String[] args) {
		int n = Integer.MAX_VALUE;
		System.out.println(Foo.isEven(n));
	}
}
 \end{lstlisting}
 
  Constant inlining is not well understood either, only 53\% of respondents answered the most simple question (CON1), inlining of a constant defined by an \texttt{int} literal, correctly. This number drops further to 33\% for inlining of string type constants (CON2), and 46\% for inlining of constants defined by expressions (inlining through constant folding, CON3). The relatively high number of correct answers for CON4 is probably misleading. Constants declared using wrapper types (\texttt{Integer} in this case) are not inlined, and we assume that many respondents answered this question accidentally correctly as they were not aware of the concept on inlining in the first place. 
 
\begin{figure}[!h]
 	\centering
 	\includegraphics[width=0.9\textwidth]{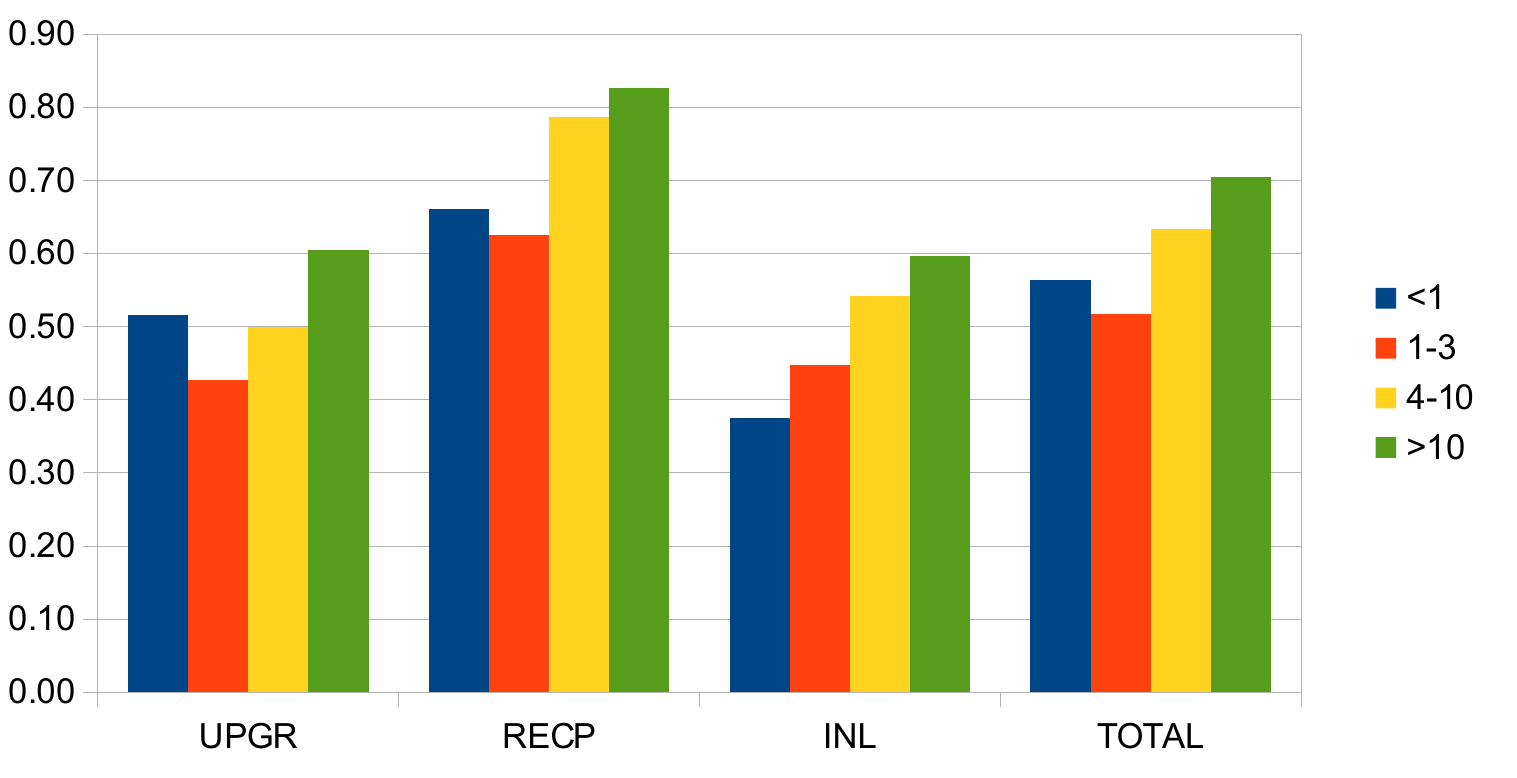}
 	\caption{Ratio of correct answers by years of java experience}
 	\label{fig:chart-yoe}
 \end{figure}

\begin{figure}[!h]
  	\centering
  	\includegraphics[width=1.0\textwidth]{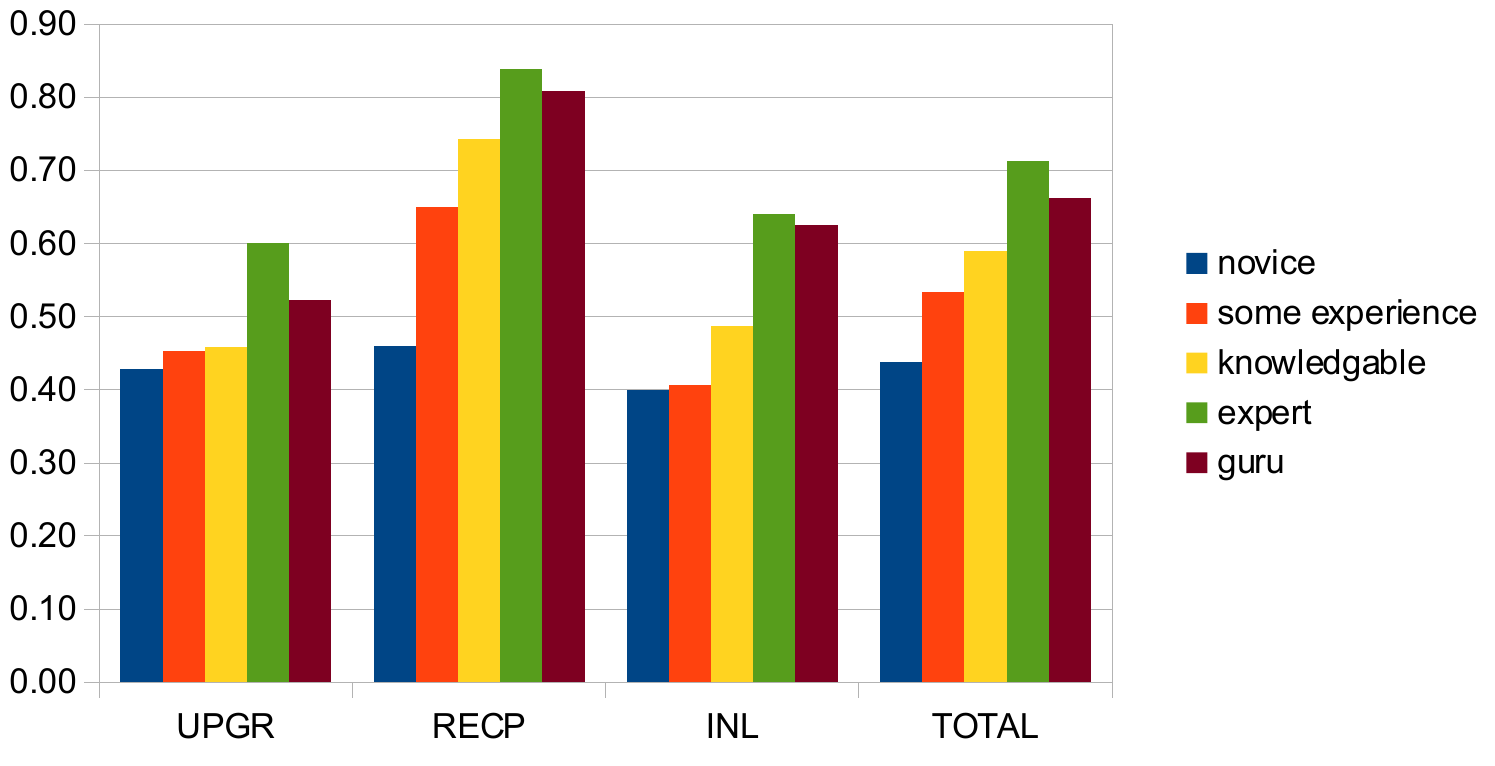}
  	\caption{Ratio of correct answers by self-assessed level of experience}
  	\label{fig:chart-loe}
  \end{figure}

Figures \ref{fig:chart-yoe} and \ref{fig:chart-loe} show the dependency of correct answers on years and level of experience with Java. Not surprisingly, answers improve with increasing experience, but not to the extent we had expected. Even (self-assessed) expert / guru users and users with more than 10 years experience of Java technology answered only 60\% of the questions in the UPGR category correctly. 
Surprisingly, the number of correct answers from participants with less than one 
year experience is slightly higher than the respective number for participants 
with 1-3 years experience. We think that this is caused by a few undergraduate 
students taught by the authors participating in the study. This particular 
cohort has some basic understanding of some of the issues involved. 
 
We further tried to work out the notion of an expert user by combining answers 
to the several contextual questions in the beginning of the survey. For this 
purpose, we defined a \textit{pro} respondent profile: a respondent with 
framework design experience, an expert or guru level of experience, at least 
four years of experience with Java technology and knowledge in at least one 
additional programming language. As OSGi is the ``killer technology'' for 
dynamic library upgrades, we also defined a second pro+osgi profile based on the 
same criteria as pro plus the requirement that the respondent must have OSGi 
experience.  For both profiles, we still had reasonable numbers of respondents 
and answers. The respective results are summarised in table 
\ref{tab:respondentprofiles}. Not surprising, pro and pro+osgi respondents 
answered more questions correctly. pro+osgi respondents do significantly better 
for the questions in the UPGR category -- they clearly better understand binary 
compatibility. However, overall the results in this category are still lower 
than expected. 

\begin{table}[h]
\caption{Question Overview}
\label{tab:respondentprofiles}
\begin{tabular}{ |l|l|l|l|l| }
\hline
questions & & all & pro & pro + osgi \\ 
\hline
\hline
\multirow{3}{*}{all}  
 & correctness &  63\% &  74\% & 78\% \\
 & answer count &  5006 &  680 & 474 \\
 & respondent count & 414 &  38 & 23 \\ \hline
 \multirow{3}{*}{UPGR} 
 & correctness &  51\% & 62\% &  69\% \\
 & answer count &  2270 & 308  & 217 \\
 & respondent count &  414 & 38 & 23 \\ \hline
\multirow{3}{*}{RECP} 
 & correctness & 76\% & 88\% &  89\% \\
 & answer count &  2263 & 308  & 217 \\
 & respondent count &  414 & 38 & 23 \\ \hline
\multirow{3}{*}{INL} 
 & correctness & 53\%  & 67\% &  70\% \\
 & answer count & 473 & 64 & 40 \\
 & respondent count & 414 & 38 & 23 \\ \hline 

\end{tabular}
\end{table}

\section{Impact Analysis}
\label{ImpactAnalysis}

\subsection{Introduction}

While the survey results demonstrate that many developers have only limited knowledge about the different kinds of compatibility rules in 
Java, it is not clear what impact this has on the quality of systems and
developer productivity. In particular, binary compatibility matters when systems
use partial upgrades. This is the case for OSGi-based technology that is used in
many developer tools and in major application servers. We therefore
\textit{suspect} that the lack of knowledge on compatibility has an impact on
both the quality of products and the productivity of developers.  In our
previous work, we found some evidence that compatibility-related problems are
common in real-world systems, and can break compatibility if libraries are
updated \cite{brokenpromises}. 

\subsection{Methodology}

To investigate the gravity of the problems caused by the lack of knowledge about compatibility rules, we focused on binary compatibility. 
The reason is that binary compatibility has a very precise definition in the Java Language
Specification \cite[ch. 13]{JL7Spec}, and violating its rules results in certain
types of linkage errors. These errors instantiate subclasses of
\texttt{java.lang.LinkageError}. This makes it easy to search for these problems
in issue tracking systems as many developers copy and paste stack traces and
error messages containing the fully qualified class name of the respective error
class. 

It is difficult to interpret absolute numbers, i.e., the number of reports for a particular error, correctly. We therefore decided to assess the number 
of linkage errors reported relative to the number of reported errors and exceptions widely considered as common. We selected two (non linkage) errors and two runtime exceptions for the baseline\footnote{All error and exception classes referenced in this section are defined in the \texttt{java.lang} package.}: 

\begin{enumerate}
\item \texttt{NullpointerException} -- this is probably the most frequently 
encountered exception  in Java, thrown when an application attempts to use 
\texttt{null} in a case where an object is required. 
\item \texttt{ClassCastException} -- thrown to indicate that an application has 
tried to cast an object to a subclass of which it is not an instance.
\item \texttt{StackOverflowError} -- thrown if an application recurses too 
deeply, usually the result of an erroneous termination condition. 
\item \texttt{OutOfMemoryError} -- thrown if the application runs out of heap 
space, e.g. if too many objects have been allocated and cannot be garbage 
collected.
\end{enumerate}

By using these four classes as a base line, we assume that these errors and 
exceptions all represent  problems that are frequently encountered by
developers. In most 
cases, they result from mistakes made by programmers, as opposed to
integration 
problems caused by the configuration of classpath settings or build scripts.  We 
have compared the frequency of problems with references to these errors and 
exceptions with the following set of linkage errors:

\begin{enumerate}
\item \texttt{NoSuchMethodError} -- thrown if an application tries to call a 
method, and this method does not exist. This usually indicates an incompatible 
change of the name or signature of the method. Many of the puzzlers, including 
most questions in the method descriptor category, cause this error to be thrown.
\item \texttt{NoSuchFieldError} -- thrown if an application tries to access a 
field, and this field does not exist. This usually indicates an incompatible 
change of the name or the type of the field. The puzzlers in the auto (un)boxing 
category cause this error to be thrown.
\item \texttt{InstantiationError} -- thrown when an application tries to 
instantiate an abstract class or interface using the \texttt{new} keyword. 
\item \texttt{IncompatibleClassChangeError} -- thrown when an incompatible class 
change has occurred to some class definition. This is the superclass of 
\texttt{NoSuchMethodError}, \texttt{NoSuchFieldError} and 
\texttt{InstantiationError}. 
\item \texttt{NoClassDefFoundError} -- thrown if an application tries to load a 
class, and this class cannot be found. This might be the result of changing the 
name of a class, or moving it into another package. 
\item \texttt{ClassFormatError} -- thrown when the JVM encounters a malformed 
class file. A common reason is that the version of the byte code is not 
supported by the JVM used.
\end{enumerate}

We have used the Google search engine to assess the frequency of errors and exceptions reported. While this methods is not very precise,
it is sufficient for us as we are mainly interested in the number of linkage
errors \textit{relative to} well-known programming errors and exceptions. We
assume that both linkage errors and the errors and exceptions used for
baselining  equally suffer from false positives and false negatives. 

We selected  three  popular open sources hosting sites according to 
\cite{PopularOpenSourceRepositories}: GitHub, SourceForge and Google Code. We 
used the \texttt{inurl} attribute in Google queries to restrict the search to 
the issues tracking system of the respective project hosting services. We also 
searched the popular Stackoverflow Q\&A site. The queries are listed in table 
\ref{tab:queries}, all queries were executed on 14 March 2014.

\begin{table}[h]
\caption{Google queries used}
\label{tab:queries}
\begin{tabular}{ |p{1.6cm}|p{9.5cm}|}
\hline
site & query\\ 
\hline \hline
github & \texttt{java.lang.NoSuchMethodError site:github.com inurl:issues} \\
sourceforge & \texttt{java.lang.NoSuchMethodError site:http://sourceforge.net/p inurl:bugs} \\
google code & \texttt{java.lang.NoSuchMethodError site:https://code.google.com/p inurl:issues} \\
stackoverflow & \texttt{java.lang.NoSuchMethodError site:stackoverflow.com} \\
\hline
\end{tabular}
\end{table}

\subsection{Results}

Table \ref{tab:errorfrequencies} shows the number of pages with references to 
the respective errors and exceptions found. The ratios differ significantly 
between different hosting sites. This is sometimes caused by few very acive 
projects with a large number of issues. For instance,  a large number of 
\texttt{NoSuchFieldErrors} appears in Google code issues. One of the Google 
code projects is \texttt{google-web-toolkit}. This project alone has 2,500 pages in its issue tracking system referencing  \texttt{NoSuchFieldError}. 

Overall, the number of linkage errors referenced is surprisingly high. In particular, \texttt{NoSuchMethodError} is in the same order of magnitude as \texttt{NullpointerException}, probably the most commonly encountered exception type in Java.  In all issues tracking systems we investigated,  \texttt{NoSuchMethodError} is more often referenced than both \texttt{StackOverflowError} and \texttt{OutOfMemoryError}. This indicates that compatibility problems already have a significant impact on the quality of products and the productivity of developers. 

\begin{table}[h]
\caption{Errors and exception reported}
\label{tab:errorfrequencies}
\begin{tabular}{ |p{4.0cm}|p{1.3cm}p{1.3cm}p{1.3cm}p{1.3cm}|}
\hline
error or exception type & github issues& google code issues& source\-forge bugs& stack\-overflow\\ 
\hline \hline
NullpointerException & 9,450 & 9,250 & 7,270 & 76,300 \\
ClassCastException & 2,030 & 6,450 & 993 & 19,600 \\
StackOverflowError & 499 & 3,630 & 237 & 51,500 \\
OutOfMemoryError & 1,080 & 4,790 & 426 & 18,000 \\ \hline 
NoSuchMethodError & 1,700 & 6,580 & 529 & 11,500\\
NoSuchFieldError & 553 & 17,400 & 108 & 2,620 \\
InstantiationError & 12 & 45 & 1 & 786 \\
IncompatibleClassChangeError & 189 & 1,770 & 51 & 2,340 \\
NoClassDefFoundError & 3,240& 2,910 & 976 & 54,700 \\
ClassFormatError & 58 & 346 & 71 & 1,600 \\
\hline
\end{tabular}
\end{table}

\section{Threats to Validity}
\label{Threats}

\subsection{Survey -- Selection bias}

The aim of this study was to find out what \textit{programmers} know about compatibility.
Indeed, most of the respondents are programmers: 282 of 338 respondents who
answered the background question (83.43\%). Note that only a very small number
of respondents are students (30 of 338 respondents who answered the background
question, 8.88\%). There is some uncertainty as 76 respondents (18.35\%) did not
answer the background questions. 

We cannot be sure how representative the population is, however, the size gives
us some confidence. We recruited respondents by advertising the survey via
the Java world\footnote{\url{
http://www.javaworld.com/article/2074970/java-library-evolution-puzzlers.html}}
portal and several Java user groups. Assuming that 
experienced developers are over-represented in these communities, the level of
knowledge about compatibility among average programmers might
actually be worse. 

\subsection{Impact Analysis -- Selection Bias}

We have only investigated projects hosted on three selected open-source hosting sites. However, we followed an independent rating \cite{PopularOpenSourceRepositories} to select hosting sites based on number of projects hosted, and the availability of a public issue tracking system. 

We have only investigated open source systems. It is likely that the same issues
appear in closed source systems, but we do not have access to a sufficiently
large number of such systems to investigate this questions. It is reasonable to
assume that closed source systems suffer from integration problems at least
the same level as open source programs, perhaps even more so as they often
represent more complex end-user products, while many open source products are
single-purpose programs or libraries not intended for use by end users, but used
as building blocks in more complex programs. 

\subsection{Impact Analysis -- False Positives}

The use of fully qualified error names makes false positives unlikely. Some
false positives could be caused by answers when respondents were confused with
similarity of Java Linkage Errors and exceptions produced by Java Reflection
API (such as
\texttt{NoSuchMethodException} vs \texttt{NoSuchMethodError}). On
the other hand, we believe that the Java Reflection API is used by experienced
users who have strong Java knowledge and thus do not tend to do such basic
mistakes.

There is a small change that linkage error class names are referenced in issues with other causes, for instance for comparison. While this is possible, it is probably very rare. Also, the other error and exception classes would suffer from this issue as well at a similar rate, so this will have no significant  effect on relative numbers.

\subsection{Impact Analysis -- False Negatives}

Since the queries try to match exact strings, it is possible that we miss
results when class names are not spelled correctly. We assume that only a few
people would type in fully qualified class names when reporting issues, instead,
copy and paste is used. Almost all issues we inspected manually  included copies
of stack traces. 
But even if there was a significant number of false negatives, this would equally affect the linkage errors and the exceptions and errors used for benchmarking. Therefore, this would have no significant effect on the relative numbers.

\section{Conclusion, Related and Future Work}
\label{Conclusion}

We have presented the results of a survey where we asked developers to solve puzzlers in order to test their knowledge on the different types of compatibility between programs and the libraries these programs use. We found that while developers have sound knowledge about the rules of source compatibility used by the compiler, even experienced developers lack knowledge about the rules of binary compatibility used by the JVM during linking.

We also demonstrated that errors which occur during linking are very common.
This seems to be consistent with a trend away from building systems from
scratch: more problems occur now at the boundaries between the actual program
and code from libraries used by the program. 

The question arises what can be done to improve the situations. 

Firstly, better education of programmers is needed. There are several  good resources available to the developer community, including some of the blog posts and presentations by Alex Buckley \cite{DifferentKindsOfCompatibility}, Joseph D. Darcy \cite{KindsOfCompatibility,JDKCompatibilityRegions}, Ian Robertson \cite{ScieneAndArtOfBackwardCompatibility} and Jim des Rivieres \cite{EvolvingJavaAPIs:2007}. We see this paper and the set of puzzlers as part of the effort.

Secondly, tools are needed to make linking smoother to facilitate library evolution. Some existing research has started to address this on different levels, including the generation of adapters to bridge "API gaps" in source code via refactoring \cite{ChowN96,Balaban:2005,Henkel:2005} and through byte code manipulation and instrumentation during class loading \cite{KellerHolzle98,JavAdaptor,Dig:2008}. Another possible approach is to change the linker itself, i.e. to build a smarter JVM with linking rules more closely aligned to the rules of source compatibility. Such a linker could support specialising return types, and would enforce the catch or re-throw rule for checked exceptions. To the best of our knowledge, this has not yet been attempted. 

Thirdly, better tools are needed to check assemblies (programs and the libraries 
they transitively depends on) for consistency. These tools could then be 
integrated into automated builds. An example is the Maven plugin that performs static byte-code
verification proposed in \cite{jezek13software} .
Another existing tool that is fairly popular amongst developers is Clirr\footnote{\url{http://sourceforge.net/projects/clirr/}}. However, we found that 
Clirr has multiple shortcomings with respect to compatibility problems 
related to declared exceptions and generic type parameters.

Finally, we notice that compatibility is complex, and that it is important to shield developers from this complexity to allow them to focus on the actual programming task. A popular method to address this is the use of semantic versioning schemes. In the Java technology space, such schemes are used in different technologies, including OSGi and Maven. Semantic versioning schemes implicitly promise contractual relationships between the providers and the consumers of APIs: if a consumer declares a requirement to use a library or module with a version within a certain range, then it is inferred that this consumer is compatible with any library or module that has a version number matching this range. In reality, this does not always work as version numbers are still assigned manually, sometimes influenced by non technical considerations such as marketing. What is therefore needed is standard tooling that can generate semantic versions complying with specifications 
such as the OSGi versioning policy \cite{OSGiVersioning} or the independent
Semantic Versioning initiative \cite{SemVer}. Automated semantic versioning has
been investigated by Bauml et al \cite{Bauml:09}. Bndtools \cite{BNDTools} is a
Java-based tool used by the OSGi community that supports semantic versioning.

\section*{Acknowledgments}

We are grateful to all developers who completed the survey. We are particularly grateful to  Jeff Friesen and Athen O'Shea from Java World, and Kon Soulianidis from the Melbourne Java user group. Without them, we would not have reached the critical number of responses. We would like to thank Alex Buckley and Hussain Al-Mutawa for their feedback on the puzzlers we used as questions in the survey.

The work was partially supported by European Regional Development Fund (ERDF), project ``NTIS - New Technologies for the Information Society'', European Centre of Excellence, CZ.1.05/1.1.00/02.0090.

\bibliographystyle{plain}
\bibliography{references}

\end{document}